\documentclass[twocolumn,showpacs,preprintnumbers,amsmath,amssymb]{revtex4}
\usepackage{graphicx}
\usepackage{dcolumn}
\usepackage{bm}
\usepackage{mathrsfs}

\begin{document}

\preprint{APS/123-QED}

\title{Black and gray spatial optical solitons with Kerr-type nonlocal nonlinearity}

\author{Shigen Ouyang}
\author{Qi Guo}
\email{guoq@scnu.edu.cn} \affiliation{Laboratory of Photonic
Information Technology, South China Normal University Guangzhou,
510631, P. R. China}

\date{\today}

\begin{abstract}
We develop one numerical method to compute black and gray solitons
with Kerr-type nonlocal nonlinearity. As two examples of nonlocal
cases, the gray soliton with exponentially decaying nonlocal
response or with Gaussian nonlocal response are discussed. For
such two nonlocal cases, the analytical form of the tails of
nonlocal gray soliton is presented and the analytical relationship
for the maximal transverse velocity of nonlocal gray soliton to
the characteristic nonlocal length is obtained.
\end{abstract}


\maketitle

\section{Introduction}
In present years spatial solitons with Kerr-type nonlocal
nonlinearity have attracted a great amount of studies. It is
indicated the nonlocality of the nonlinearity ensures the
existence of stable multidimensional solitons\cite{1,2,3,4}. The
nonlocal bright spatial solitons have been experimentally observed
in nematic liquid crystal\cite{5,6,7} and in the lead
glass\cite{8}. The propagation and interaction properties of
nonlocal bright/black solitons are greatly different from that of
local solitons. For example, the dependent functions of the beam
power and phase constant on the beam width for nonlocal bright
solitons are very different from those for local bright
solitons\cite{9,10,11}; There exists higher order Hermite-Gaussian
like (1+1) dimensional bright nonlocal solitons\cite{9,10,12} and
Laguerre-Gaussian like (1+2) dimensional bright nonlocal
solitons\cite{9,11,12,13,14,15}; Bright nonlocal solitons with
$\pi$ phase difference attract rather repel each
other\cite{12,16,17,18}; Nonlocal black solitons can attract each
other and two black solitons can form a bound
state\cite{19,20,21}. The nonlocal gray solitons also can form a
bound state\cite{24}.

In this paper we present a numerical method to compute black and
gray soliton solutions with Kerr-type nonlocal nonlinearity. This
numerical method is a generalized version of the so called
spectral renormalization method presented by M. J. Ablowitz and Z.
H. Musslimani\cite{22} with which they compute the bright soliton
solutions. Our numerical method can be easily generalized to
compute a large family of gray soliton solutions with other type
of nonlinearity. It is worth to note that Yaroslav V. Kartashov
and Lluis Torner\cite{24} has rather successfully studied the
nonlocal gray soliton with an exponential decaying nonlocal
response in a different manner. They has revealed that the gray
soliton velocity depends on the nonlocality degree and pointed out
the maximal velocity of gray soliton monotonically decreases with
the degree of nonlocality\cite{24}. In our paper, besides of the
nonlocal case of an exponential decaying nonlocal response, the
nonlocal case of Gaussian nonlocal response is also discussed. The
analytical form of the tails of nonlocal gray soliton is
presented. It is indicated that nonlocal gray solitons can have
exponentially decaying tails or have exponentially decaying
oscillatory tails. The analytical relationship for the maximal
transverse velocity of nonlocal gray soliton to the characteristic
nonlocal length is presented. It is indicated when the
characteristic nonlocal length is less than some critical value,
such maximal transverse velocity is a constant and equal to that
of local gray soliton, otherwise such maximal velocity will
decrease with the increasing of the characteristic nonlocal
length. That mostly agrees with the results of reference[24] with
slight but not trivial exception. It was said in reference[24]
that such maximal velocity monotonically decreases with the
characteristic nonlocal length. But as will be shown in our paper
the maximal velocity does not vary with the characteristic
nonlocal length when the characteristic nonlocal length is less
than some critical value. Our paper poses a problem for such
inconsistence.

\section{Numerical method to compute nonlocal gray solitons}
The propagation of an (1+1) dimensional optical beam in Kerr-type
nonlocal self-defocusing media can be described by this following
(1+1) dimensional nonlocal nonlinear Sch\"{o}dinger
equation(NNLSE)\cite{1,2,3,4,5,6,7,8,9,10,11,12,13,14,15,16,17,18,19,20,21}
\begin{equation}\label{1-1}
i{\partial u\over\partial z}+\frac{1}{2}{\partial^2 u\over\partial
x^2}-u\int R(x-\xi)|u(\xi,z)|^2d\xi=0,
\end{equation}
where $u(x,z)$ is the complex amplitude envelop of the light beam,
$|u(x,z)|^2$ is the light intensity, $x$ and $z$ are transverse
and longitude coordinates respectively, $R(x)$, ($\int R(x)dx=1$)
is the real symmetric nonlocal response function, and $
n(x,z)=-\int R(x-\xi)|u(\xi,z)|^2d\xi$ is the light-induced
perturbed refractive index. Note that not stated otherwise all
integrals in this paper will extend over the whole x axis. When
$R(x)=\delta(x)$, equation~(\ref{1-1}) will reduce to the local
nonlinear Sch\"{o}dinger equation(NLSE)
\begin{eqnarray}\label{1-2}
i{\partial u\over\partial z}+\frac{1}{2}{\partial^2 u\over\partial
x^2}-|u|^2u=0,
\end{eqnarray}
which has black and gray soliton solutions\cite{23}
\begin{subequations}
\begin{eqnarray}
u(x,z)&&=\psi(x)e^{i[\beta z+\phi(x)]}\\
\psi(x)&&=\eta[1-B^2\rm{sech}^2(\eta B x)]^{1/2}\\
\beta&&=-{{1}\over{2}}\eta^2(3-B^2)\\
\phi(x)&&=\eta\sqrt{1-B^2}x+\arctan\left[{{B\tanh(\eta B
x)}\over{\sqrt{1-B^2}}}\right]
\end{eqnarray}
\end{subequations}
It is easy to prove that
$\psi(x)\xrightarrow{x\rightarrow+\infty}\eta$ and
$\phi^\prime(x)\xrightarrow{x\rightarrow+\infty}\eta\sqrt{1-B^2}\equiv\mu$.
So we have $\beta=-({{\mu^2}\over{2}}+\eta^2)$ which, as will be
shown, also applies to nonlocal gray soliton.

In this paper we numerically compute the black and gray soliton
solutions of NNLSE~(\ref{1-1}) that take these following form
\begin{subequations}
\begin{eqnarray}
&&u(x,z)=\psi(x)e^{i\beta z+i\phi(x)}\label{1-3}\\
&&\psi^*(x)=\psi(x)~~~~~\lim_{x\rightarrow+\infty}\psi(x)=\eta>0\label{1-3-1}\\
&&\beta^*=\beta~~~~~~~~~~~~~~~\phi^*(x)=\phi(x)
\end{eqnarray}
\end{subequations}
Substitution of Eq.~(\ref{1-3}) into (\ref{1-1}), we have an
equation for $\psi(x)$ and $\phi(x)$. Since both the real and
imaginary part of such resulting equation must vanish, we have
\begin{eqnarray}\label{1-4}
-\beta\psi+{{1}\over{2}}\psi^{\prime\prime}-{{1}\over{2}}(\phi^\prime)^2\psi-\psi\int
R(x-\xi)\psi^2(\xi)d\xi=0
\end{eqnarray}
\begin{eqnarray}\label{1-5}
2\phi^\prime\psi^\prime+\phi^{\prime\prime}\psi=0
\end{eqnarray}
From Eq.~(\ref{1-5}),we have
${{d}\over{dx}}(\psi^2\phi^\prime)=0$, which results in
$\psi^2\phi^\prime=\rm{const}$. We set this constant as
$\mu\eta^2$, where $\mu^*=\mu$ is another constant. So we have
\begin{eqnarray}\label{1-6}
\phi^\prime={{\mu\eta^2}\over{\psi^2}}.
\end{eqnarray}
Since $\psi(x)\xrightarrow{x\rightarrow+\infty}\eta$ as assumed in
Eq.~(\ref{1-3-1}), from Eq.~(\ref{1-6}), we have
$\phi^\prime(x)\xrightarrow{x\rightarrow+\infty}\mu$, which in
turn results in $\phi(x)\xrightarrow{x\rightarrow+\infty}c+\mu x$,
where $c$ is a constant. The phase jump through the soliton is
defined as $2c$. Substituting Eq.~(\ref{1-6}) into (\ref{1-4}), we
get
\begin{eqnarray}\label{1-8}
-\beta\psi+{{1}\over{2}}\psi^{\prime\prime}-{{\mu^2\eta^4}\over{2\psi^3}}-\psi\int
R(x-\xi)\psi^2(\xi)d\xi=0.
\end{eqnarray}
Since $\psi(x)\xrightarrow{x\rightarrow+\infty}\eta$ and $\int
R(x)dx=1$, from Eq.~(\ref{1-8}), when $x\rightarrow+\infty$, we
have $-\beta\eta-{{\mu^2\eta}\over{2}}-\eta^3=0$ that leads to
\begin{eqnarray}
\beta=-\left({{\mu^2}\over{2}}+\eta^2\right),
\end{eqnarray}
and
\begin{eqnarray}\label{1-9}
{{\mu^2+2\eta^2}\over{2}}\psi+{{\psi^{\prime\prime}}\over{2}}-{{\mu^2\eta^4}\over{2\psi^3}}-\psi\int
R(x-\xi)\psi^2(\xi)d\xi=0.\nonumber\\
\end{eqnarray}
So the gray soliton takes a form of
$u(x,z)=\psi(x)\exp[-i({{\mu^2}\over{2}}+\eta^2)z+i\phi(x)]$,
where $\psi(x),\phi(x)$ satisfy Eqs.~(\ref{1-9}) and (\ref{1-6}).
It is easy to prove that the Galilean transformed solution
\begin{eqnarray}\label{1-16}
u_1(x,z)&&\equiv u(x+\mu z,z)e^{i(-\mu
x-{{1}\over{2}}\mu^2z)}\nonumber\\
&&=\psi(x+\mu z)e^{i[-\eta^2z+\phi(x+\mu z)-\mu(x+\mu
z)]}\nonumber\\
&&=\psi_1(x+\mu z)e^{-i\eta^2z},
\end{eqnarray}
where $\psi_1(x+\mu z)\equiv\psi(x+\mu z)e^{i[\phi(x+\mu
z)-\mu(x+\mu z)]}$, also satisfies the NNLSE~(\ref{1-1}). Since
$|u_1(x,z)|^2=|\psi_1(x+\mu z)|^2=\psi^2(x+\mu z)$, the gray
soliton moves in velocity $-\mu$ with respect to the coordinate
system. On the other hand, since
$\psi(x)\xrightarrow{x\rightarrow+\infty}\eta$ and
$\phi(x)\xrightarrow{x\rightarrow+\infty}c+\mu x$, from
Eq.~(\ref{1-16}), we have
$u_1(x,z)\xrightarrow{x\rightarrow+\infty}\eta e^{i(c-\eta^2z)}$,
which implies the background of gray soliton is at rest with
respect to the coordinate system. So the soliton $u_1(x,z)$ moves
in transverse velocity $-\mu$ with respect to the background
intensity of gray soliton. In this paper we simply refer $-\mu$ as
the transverse velocity of the soliton.

Assuming $\psi(x)\xrightarrow{x\rightarrow-\infty}\eta^\prime$,
from Eq.~(\ref{1-9}), we have $\eta^\prime=\pm\eta$. When
$\mu\neq0$,the term ${{\mu^2\eta^4}\over{2\psi^3}}$ requires
$\psi(x)\neq0$, which in turn requires $\eta^\prime=\eta$. So the
case $\mu\neq0$ corresponds to the gray solitons and
$\psi(-x)=\psi(x)$, where we have set $x=0$ be the symmetric
center.

We study the case $\mu\neq0$ in this section and leave the case
$\mu=0$ to be discussed in the next section. Let
\begin{eqnarray}\label{1-10}
\psi(x)=\eta-\chi(x),
\end{eqnarray}
where $\chi(-x)=\chi(x)$ and
$\chi(x)\xrightarrow{x\rightarrow\pm\infty}0$. Then
Eq.~(\ref{1-9}) turns into
\begin{eqnarray}\label{1-11}
&&-{{\mu^2}\over{2}}\chi-{{1}\over{2}}\chi^{\prime\prime}
+{{\mu^2\eta}\over{2}}\cdot{{3\eta\chi^2-3\eta^2\chi-\chi^3}\over{(\eta-\chi)^3}}\nonumber\\
&&~-(\eta-\chi)\int R(x-\xi)[\chi^2(\xi)-2\eta\chi(\xi)]d\xi=0.
\end{eqnarray}
We discrete the function $\chi(x)$ in
$\chi_{j}=\chi\big(-h+(j-1)\triangle x\big)$, where $-h<x<h$ is
the sample window, $\triangle x$ is the sample step and $1\leq
j\leq n={{2h}\over{\triangle x}}+1$. Define the discrete Fourier
transform (DFT) $\mathscr{F}$ by
\begin{subequations}
\begin{eqnarray}
&&\widetilde{\chi}_{j}=\mathscr{F}[\chi]_{j}=\sum_{k=1}^n\rm{F}_{jk}\chi_{k},\\
&&\chi_{j}=\mathscr{F}^{-1}[\widetilde{\chi}]_{j}=\sum_{k=1}^n\rm{F}^*_{jk}\widetilde{\chi}_{k},
\end{eqnarray}
\end{subequations}
where
$\rm{F}_{jk}={{1}\over{\sqrt{n}}}\exp[i{{2\pi}\over{n}}(j-1)(k-1)]$.
Performing the DFT on Eq.~(\ref{1-11}), we have
\begin{eqnarray}\label{1-12}
&&-{{\mu^2}\over{2}}\widetilde{\chi}_j+{{\Omega_j}\over{2}}\widetilde{\chi}_j
+{{\mu^2\eta}\over{2}}\mathscr{F}\left[{{3\eta\chi^2-3\eta^2\chi-\chi^3}\over{(\eta-\chi)^3}}\right]_j\nonumber\\
&&-\mathscr{F}\left[(\eta-\chi)\int
R(x-\xi)[\chi^2(\xi)-2\eta\chi(\xi)]d\xi\right]_j=0,\nonumber\\
\end{eqnarray}
where
$\Omega_{j}=\left({{2\sin[{{\pi}\over{n}}(j-1)]}\over{\triangle
x}}\right)^2$. Let
\begin{eqnarray}
\chi(x)=\lambda\theta(x),
\end{eqnarray}
we have
\begin{eqnarray}\label{1-13}
&&-{{\mu^2}\over{2}}\widetilde{\theta}_j+{{\Omega_j}\over{2}}\widetilde{\theta}_j
+{{\mu^2\eta}\over{2}}\mathscr{F}\left[{{3\lambda\eta\theta^2-3\eta^2\theta-\lambda^2\theta^3}
\over{(\eta-\lambda\theta)^3}}\right]_j\nonumber\\
&&-\mathscr{F}\left[(\eta-\lambda\theta)\int
R(x-\xi)[\lambda\theta^2(\xi)-2\eta\theta(\xi)]d\xi\right]_j=0.\nonumber\\
\end{eqnarray}
Projecting Eq.~(\ref{1-13}) onto $\widetilde{\theta}$, we obtain
an equation for $\lambda$
\begin{eqnarray}\label{1-14}
&&~~\mathscr{A}_{\theta}(\lambda)\nonumber\\
&&\equiv\sum_{j=1}^n\widetilde{\theta}_j^*
\bigg\{-{{\mu^2}\over{2}}\widetilde{\theta}_j+{{\Omega_j}\over{2}}\widetilde{\theta}_j\nonumber\\
&&+{{\mu^2\eta}\over{2}}\mathscr{F}\left[{{3\lambda\eta\theta^2-3\eta^2\theta-\lambda^2\theta^3}
\over{(\eta-\lambda\theta)^3}}\right]_j\nonumber\\
&&-\mathscr{F}\left[(\eta-\lambda\theta)\int
R(x-\xi)[\lambda\theta^2(\xi)-2\eta\theta(\xi)]d\xi\right]_j\bigg\}\nonumber\\
&&=0.
\end{eqnarray}
On the other hand, from Eq.~(\ref{1-13}), we get
\begin{eqnarray}\label{1-15}
\widetilde{\theta}_j&&={{r\widetilde{\theta}_j+{{\mu^2}\over{2}}\widetilde{\theta}_j-{{\mu^2\eta}\over{2}}\mathscr{F}\left[{{3\lambda\eta\theta^2-3\eta^2\theta-\lambda^2\theta^3}
\over{(\eta-\lambda\theta)^3}}\right]_j}\over{r+{{\Omega_j}\over{2}}}}\nonumber\\
&&+{{\mathscr{F}\left[(\eta-\lambda\theta)\int
R(x-\xi)[\lambda\theta^2(\xi)-2\eta\theta(\xi)]d\xi\right]_j}\over{r+{{\Omega_j}\over{2}}}}\nonumber\\
&&\equiv\mathscr{D}_{\lambda}[\theta]_j
\end{eqnarray}
where $r$ is a positive constant.

We use Eqs.~(\ref{1-14}) and (\ref{1-15}) to iteratively compute
gray soliton solutions. For an initial function $\theta_1(x)$,
e.g. a Gaussian function, from Eq.~(\ref{1-14}), we get
$\lambda_1$ which satisfies the equation
$\mathscr{A}_{\theta_1}(\lambda_1)=0$. Then from Eq.~(\ref{1-15})
we get function
$\widetilde{\theta}_2=\mathscr{D}_{\lambda_1}[\theta_1]$. For
$m\geq1$, we get the iteration scheme
$\mathscr{A}_{\theta_m}(\lambda_m)=0,
\widetilde{\theta}_{m+1}=\mathscr{D}_{\lambda_m}[\theta_m]$.
Perform the iteration until the convergence is achieved. Then we
get $\psi(x)=\eta-\lambda\theta(x)$ and
$\phi(x)=\int_0^x{{\mu\eta^2}\over{\psi^2(\xi)}}d\xi$.

\begin{figure}
\centering
\includegraphics[totalheight=2.8in]{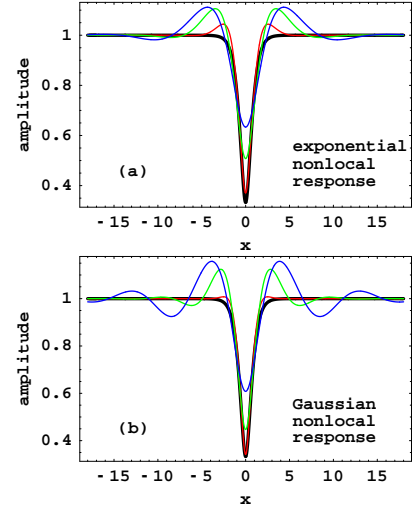}
\caption{\label{fig01}Amplitudes of gray soliton solutions with
exponential nonlocal response (a) or with Gaussian nonlocal
response (b). The parameters used are $\eta=1, \mu=1/3$ and black
line corresponds to local case $w=0$, red line $w=1$, green line
$w=3$ and blue line $w=5$ both for (a) and (b).}
\end{figure}

\begin{figure}
\centering
\includegraphics[totalheight=1.8in]{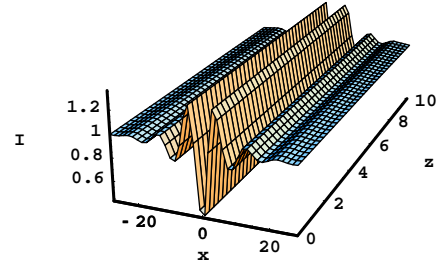}
\caption{\label{fig03}Numerical simulation of the gray soliton
with Gaussian nonlocal response and with parameters $w=5, \eta=1,
\mu=1/3$.}
\end{figure}

\begin{figure}
\centering
\includegraphics[totalheight=1.5in]{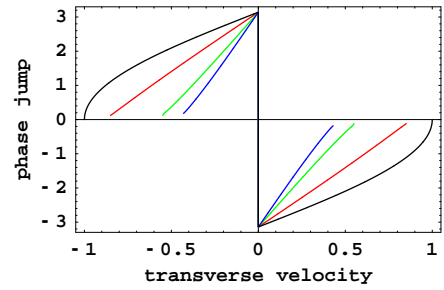}
\caption{\label{fig04}Phase jump through the nonlocal gray soliton
with an exponential nonlocal response. Black line is the local
case $w=0$; Red line $w=1$; Green line $w=3$; Blue line $w=5$.}
\end{figure}

\begin{figure}
\centering
\includegraphics[totalheight=2.7in]{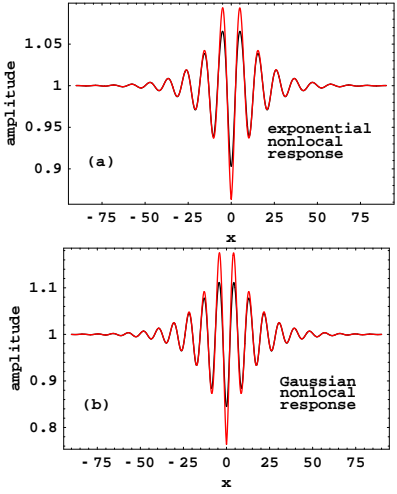}
\caption{\label{fig02}(a) Black line is the amplitude of gray
soliton solution with exponential nonlocal response and with
parameters $\eta=1,\mu=0.43,w=5$, and the red line is the fitting
function $1-0.137\exp(-0.0753|x|)\cos(0.596|x|+0.055)$; (b) Gray
soliton solution with Gaussian nonlocal response with parameters
$\eta=1,\mu=0.4,w=5$ and the fitting function
$1-0.238\exp(-0.0733|x|)\cos(0.713|x|+0.107)$.}
\end{figure}

As an example, we consider this following nonlocal
case\cite{1,2,3,4,5,6,7,10,11,16,18,21,24}
\begin{eqnarray}
n-w^2{{d^2n}\over{dx^2}}=-|u|^2,
\end{eqnarray}
which results in
$n(x)=-\int{{1}\over{2w}}\exp(-{{|x-\xi|}\over{w}})|u(\xi)|^2d\xi$,
where $R(x)={{1}\over{2w}}\exp(-{{|x|}\over{w}})$ is the
exponential decaying nonlocal response function. In another
example, we consider the Gaussian nonlocal
response\cite{2,3,9,10,11,14,15,17,18}
$R(x)={{1}\over{w\sqrt{\pi}}}\exp\left(-{{x^2}\over{w^2}}\right)$.
For such two nonlocal cases, $w$ is referred as the characteristic
nonlocal length. In Fig.~(\ref{fig01}) some nonlocal gray soliton
solutions are shown. The numerical simulation result in
Fig.~(\ref{fig03}) indicates the numerical nonlocal gray soliton
solution obtained can describe the soliton state very well. The
phase jump through the nonlocal gray soliton is shown in
Fig.~(\ref{fig04}). As shown by Fig.~(\ref{fig02}) the nonlocal
gray solitons can have exponentially decaying oscillatory tails
for both the exponential nonlocal response case and the Gaussian
nonlocal response case.

Now we investigate the form of the decaying tails of gray
solitons. In case $|\chi(x)|\ll\eta$, to the leading order
Eq.~(\ref{1-11}) can be linearized to
\begin{eqnarray}\label{1-17}
-2\mu^2\chi-{{1}\over{2}}\chi^{\prime\prime}+2\eta^2\int
R(x-\xi)\chi(\xi)d\xi=0.
\end{eqnarray}
Since Eq.~(\ref{1-17}) is linear for $\chi(x)$, the superposition
theorem applies.

For the exponential nonlocal response
$R(x)={{1}\over{2w}}\exp(-{{|x|}\over{w}})$, Eq.~(\ref{1-17})
reduces to two coupled equations
\begin{subequations}\label{1-18}
\begin{eqnarray}
&&-2\mu^2\chi-{{1}\over{2}}\chi^{\prime\prime}+2\eta^2f(x)=0,\label{1-18a}\\
&&f(x)-w^2f^{\prime\prime}(x)=\chi(x).\label{1-18b}
\end{eqnarray}
\end{subequations}
The solution of (\ref{1-18}) can be assumed as
\begin{eqnarray}\label{1-18c}
\chi(x)=\alpha\exp(\lambda x).
\end{eqnarray}
Substituting (\ref{1-18c}) into (\ref{1-18}), we have
\begin{eqnarray}
\chi^{\prime\prime}=4\left({{\eta^2}\over{1-\lambda^2w^2}}-\mu^2\right)\chi
\end{eqnarray}
The eigenvalue problem of the above equation provides an equation
for $\lambda$
\begin{eqnarray}
\lambda^2=4\left({{\eta^2}\over{1-\lambda^2w^2}}-\mu^2\right),
\end{eqnarray}
which results in
\begin{eqnarray}\label{1-19}
\lambda^2={{1-4\mu^2w^2-\sqrt{1+8\mu^2w^2+16\mu^4w^4-16\eta^2w^2}}\over{2w^2}}.
\end{eqnarray}
When $w\rightarrow0$, from Eq.~(\ref{1-19}) to the leading order
we have
\begin{eqnarray}
\lambda^2=4(\eta^2-\mu^2)(1+4\eta^2w^2)
\end{eqnarray}
The roots of Eq.~(\ref{1-19}) are
\begin{eqnarray}
\lambda=\pm(\lambda_1+i\lambda_2),
\end{eqnarray}
where
\begin{eqnarray}
&&\lambda_1=\sqrt{{{1-4w^2\mu^2+4w\sqrt{\eta^2-\mu^2}}\over{4w^2}}},\\
&&\lambda_2=\sqrt{{{-1+4w^2\mu^2+4w\sqrt{\eta^2-\mu^2}}\over{4w^2}}}.
\end{eqnarray}
For $w=5, \eta=1, \mu=0.43$, we have
$\lambda_1=0.0753,\lambda_2=0.596$ which are used by the fitting
function in Fig.~(\ref{fig02}).

To obtain exponentially decaying tails, $\lambda$ must be a real
number and we have
\begin{subequations}\label{1-20}
\begin{eqnarray}
1-4w^2\mu^2+4w\sqrt{\eta^2-\mu^2}&&>0,\\
-1+4w^2\mu^2+4w\sqrt{\eta^2-\mu^2}&&\leq0,
\end{eqnarray}
\end{subequations}
which in turn result into
\begin{subequations}\label{1-21}
\begin{eqnarray}
0\leq&&\mu^2\leq\eta^2~~~\rm{for}~~~~~~~~~~~w\leq{{1}\over{4\eta}},\\
{{4w\eta-1}\over{4w^2}}\leq&&\mu^2\leq\eta^2~~~\rm{for}~~~{{1}\over{4\eta}}\leq
w\leq{{1}\over{2\eta}}.
\end{eqnarray}
\end{subequations}
On the other hand when
\begin{subequations}\label{1-22}
\begin{eqnarray}
1-4w^2\mu^2+4w\sqrt{\eta^2-\mu^2}&&>0,\\
-1+4w^2\mu^2+4w\sqrt{\eta^2-\mu^2}&&>0,
\end{eqnarray}
\end{subequations}
$\lambda$ is a complex number and exponentially decaying
oscillatory tails are obtained. From inequalities~(\ref{1-22}), we
have
\begin{eqnarray}
0\leq\mu^2<{{4w\eta-1}\over{4w^2}}\leq\eta^2~~~~~~~~\rm{for}~~~~~~~~w>{{1}\over{4\eta}}.
\end{eqnarray}

\begin{figure}
\centering
\includegraphics[totalheight=1.5in]{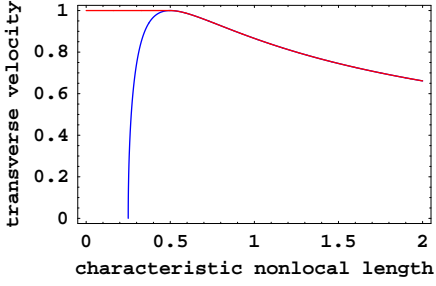}
\caption{\label{fig06}The area under the red line is the parameter
space for gray soliton with exponential decaying nonlocal response
when $\eta=1$. The left area of the blue line corresponds to gray
solitons with exponentially decaying tails; the right area
corresponds to gray solitons with exponentially decaying
oscillatory tails.}
\end{figure}

In conclusion, when $w\leq{{1}\over{2\eta}}$ the maximal
transverse velocity does not vary with the characteristic nonlocal
length $w$ and is equal to $\eta$ the maximal velocity of local
gray soliton; When $w>{{1}\over{2\eta}}$ the maximal transverse
velocity is $\sqrt{{{4w\eta-1}\over{4w^2}}}$. The maximal
transverse velocity will decrease when $w>{{1}\over{2\eta}}$ with
the increasing of the nonlocal length $w$. Such results are shown
in Fig.~(\ref{fig06}). It is worth to note that there is slight
but not trivial inconsistence between our results and the results
of reference[24]. As shown in figure 4(a) of reference[24] and
pointed out in reference[24], the maximal velocity monotonically
decreases with the nonlocality degree (the characteristic nonlocal
length in this paper). But as has been pointed out by our paper
and shown in figure (\ref{fig06}) in this paper, the maximal
velocity does not vary with the characteristic nonlocal length $w$
when $w\leq{{1}\over{2\eta}}$.

On the form of the tail of the nonlocal gray soliton we arrive at,
for $w\leq{{1}\over{4\eta}}$, the gray solitons always have
exponentially decaying tails; for $w>{{1}\over{2\eta}}$, the gray
solitons always have exponentially decaying oscillatory tails; for
${{1}\over{4\eta}}<w\leq{{1}\over{2\eta}}$, the gray solitons can
have exponentially decaying oscillatory tails when
$0\leq\mu^2<{{4w\eta-1}\over{4w^2}}$ or have exponentially
decaying tails when ${{4w\eta-1}\over{4w^2}}\leq\mu^2\leq\eta^2$.

Now we consider the nonlocal case with a Gaussian nonlocal
response $R(x)={{1}\over{w\sqrt{\pi}}}\exp(-{{x^2}\over{w^2}})$.
Here we only discuss the exponentially decaying oscillatory tails.
However the exponentially decaying tails can be discussed in a
similar way. For the exponentially decaying oscillatory tails
\begin{eqnarray}
\chi(x)=\exp(-\lambda x)\cos(\kappa x),
\end{eqnarray}
we have
\begin{eqnarray}
f(x)&&=\int_{-\infty}^{+\infty}{{1}\over{w\sqrt{\pi}}}\exp\left[-{{(x-\xi)^2}\over{w^2}}\right]\chi(\xi)d\xi\nonumber\\
&&=A\chi(x)+B\chi^\prime(x),
\end{eqnarray}
where
\begin{eqnarray}
&&A=e^{{w^2(\lambda^2-\kappa^2)}\over{4}}\left(\cos{{w^2\lambda\kappa}\over{2}}-{{\lambda}\over{\kappa}}\sin{{w^2\lambda\kappa}\over{2}}\right),\\
&&B=-{{1}\over{\kappa}}e^{{w^2(\lambda^2-\kappa^2)}\over{4}}\sin{{w^2\lambda\kappa}\over{2}}.
\end{eqnarray}
So Eq.~(\ref{1-17}) turns into
\begin{eqnarray}
-{{1}\over{2}}\chi^{\prime\prime}+2\eta^2B\chi^\prime+(2\eta^2A-2\mu^2)\chi=0.
\end{eqnarray}
The eigenvalue problem of the above equation provides two coupled
equations for $\lambda$ and $\kappa$
\begin{subequations}\label{1-23}
\begin{eqnarray}
&&\lambda=-2\eta^2B,\\
&&\kappa=\sqrt{-2(2\eta^2A-2\mu^2)-4\eta^4B^2}
\end{eqnarray}
\end{subequations}
For example, when $w=5, \eta=1, \mu=0.4$, from (\ref{1-23}) we get
$\lambda=0.0733, \kappa=0.713$ which are used by the fitting
function in Fig.~(\ref{fig02}).

\begin{figure}
\centering
\includegraphics[totalheight=1.5in]{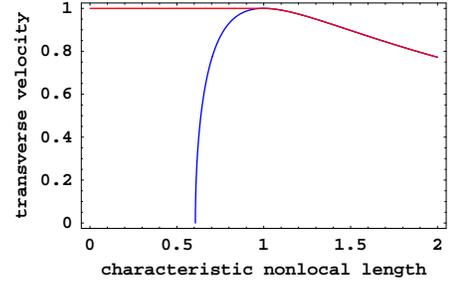}
\caption{\label{fig07}The area under the red line is the parameter
space for gray soliton with Gaussian nonlocal response when
$\eta=1$. The left area of the blue line corresponds to gray
solitons with exponentially decaying tails; the right area
corresponds to gray solitons with exponentially decaying
oscillatory tails.}
\end{figure}

In a similar way of deducing the parameter space in the
exponentially decaying nonlocal case, we also get the parameter
space in the Gaussian nonlocal case. When $w<1/\eta$, the maximal
transverse velocity is a constant equal to $\eta$; when
$w>1/\eta$, the maximal velocity is equal to
$\sqrt{{1+\ln(\eta^2w^2)}\over{w^2}}$. On the other hand when
$w>1/\eta$, the gray soliton always has exponentially decaying
oscillatory tail; When $w\leq1/\sqrt{\rm{e}}\eta$, the gray
soliton always has exponentially decaying tail; when
$1/\sqrt{\rm{e}}\eta<w<1/\eta$, the gray soliton can have
exponentially decaying tail for
$\eta\geq\mu>\sqrt{{1+\ln(\eta^2w^2)}\over{w^2}}$ or exponentially
decaying oscillatory tail for
$\mu<\sqrt{{1+\ln(\eta^2w^2)}\over{w^2}}$. These results are shown
in Fig.~(\ref{fig07}).

\section{Numerical method to compute nonlocal black soliton solutions}
In the case $\mu=0$, Eq.~(\ref{1-9}) reduces to
\begin{eqnarray}\label{2-1}
\eta^2\psi+{{\psi^{\prime\prime}}\over{2}}-\psi\int
R(x-\xi)\psi^2(\xi)d\xi=0.
\end{eqnarray}
We consider the nonlocal black soliton solutions with properties
\begin{eqnarray}\label{2-1-1}
\psi(-x)=-\psi(x)~~~~\psi(x)\xrightarrow{x\rightarrow\pm\infty}\pm\eta~~~~\psi(0)=0
\end{eqnarray}
Let
\begin{eqnarray}\label{2-1-2}
\chi(x)={{d\psi(x)}\over{dx}}~~~\rm{i.e.}~~~\psi(x)=\int_0^x\chi(t)dt
\end{eqnarray}
So $\chi(-x)=\chi(x)$ and
$\chi(x)\xrightarrow{x\rightarrow\pm\infty}0$. Acting
${{d}\over{dx}}$ on Eq.~(\ref{2-1}), we obtain
\begin{eqnarray}\label{2-1-3}
&&\eta^2\chi+{{1}\over{2}}\chi^{\prime\prime}-\chi\int
R(x-\xi)\psi^2(\xi)d\xi\nonumber\\
&&-2\psi\int R(x-\xi)\psi(\xi)\chi(\xi)d\xi=0
\end{eqnarray}
Let
\begin{eqnarray}
\chi(x)=\lambda\theta(x),
\end{eqnarray}
and substitute it into Eq.~(\ref{2-1-3}), we have
\begin{eqnarray}\label{2-1-4}
&&\eta^2\theta+{{1}\over{2}}\theta^{\prime\prime}-\lambda^2\theta\int
R(x-\xi)\psi_1^2(\xi)d\xi\nonumber\\
&&-2\lambda^2\psi_1\int R(x-\xi)\psi_1(\xi)\theta(\xi)d\xi=0,
\end{eqnarray}
Where $\psi_1(x)\equiv\int_0^x\theta(t)dt$. Performing DFT on
Eq.~(\ref{2-1-4}), we have
\begin{eqnarray}\label{2-1-5}
&&\eta^2\widetilde{\theta}_j-{{\Omega_j}\over{2}}\widetilde{\theta}_j-\lambda^2\mathscr{F}\left[\theta\int
R(x-\xi)\psi_1^2(\xi)d\xi\right]_j\nonumber\\
&&-\lambda^2\mathscr{F}\left[2\psi_1\int
R(x-\xi)\psi_1(\xi)\theta(\xi)d\xi\right]_j=0.
\end{eqnarray}
\begin{figure}
\centering
\includegraphics[totalheight=1.3in]{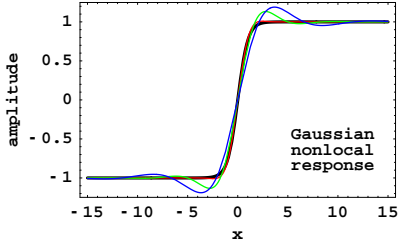}
\caption{\label{fig05}Amplitudes of nonlocal black soliton
solutions with Gaussian nonlocal response. The parameters used are
$\eta=1$ and black line $w=0$, red line $w=1$, green line $w=3$
and blue line $w=5$.}
\end{figure}
Projecting Eq.~(\ref{2-1-5}) onto $\widetilde{\theta}$, we obtain
an equation for $\lambda$
\begin{eqnarray}\label{2-1-6}
&&~~\mathscr{A}_{\theta}(\lambda)\nonumber\\
&&\equiv\sum_{j=1}^n\widetilde{\theta}^*_j\bigg\{\eta^2\widetilde{\theta}_j
-\lambda^2\mathscr{F}\left[\theta\int
R(x-\xi)\psi_1^2(\xi)d\xi\right]_j\nonumber\\
&&-{{\Omega_j}\over{2}}\widetilde{\theta}_j-\lambda^2\mathscr{F}\left[2\psi_1\int
R(x-\xi)\psi_1(\xi)\theta(\xi)d\xi\right]_j\bigg\}=0\nonumber\\
\end{eqnarray}
On the other hand, from Eq.~(\ref{2-1-5}) we have
\begin{eqnarray}\label{2-1-7}
\widetilde{\theta}_j&&={{r\widetilde{\theta}_j+\eta^2\widetilde{\theta}_j-\lambda^2\mathscr{F}\left[\theta\int
R(x-\xi)\psi_1^2(\xi)d\xi\right]_j}\over{r+{{\Omega_j}\over{2}}}}\nonumber\\
&&~~-{{\lambda^2\mathscr{F}\left[2\psi_1\int
R(x-\xi)\psi_1(\xi)\theta(\xi)d\xi\right]_j}\over{r+{{\Omega_j}\over{2}}}}\nonumber\\
&&\equiv\mathscr{D}_{\lambda}[\theta]_j
\end{eqnarray}

The iteration scheme is $\mathscr{A}_{\theta_m}(\lambda_m)=0,
\widetilde{\theta}_{m+1}=\mathscr{D}_{\lambda_m}[\theta_m]$. Some
black nonlocal soliton solutions with Gaussian nonlocal response
are shown in Fig.~(\ref{fig05}).

\section{conclusion}
One numerical method to compute black and gray soliton solutions
with Kerr-type nonlocal nonlinearity is developed. As two examples
of nonlocal cases, the gray soliton with exponentially decaying
nonlocal response or with Gaussian nonlocal response are
discussed. The analytical form of the tails of nonlocal gray
soliton is presented. It is indicated that nonlocal gray solitons
can have exponentially decaying tails or have exponentially
decaying oscillatory tails. The analytical relationship for the
maximal transverse velocity of nonlocal gray soliton to the
characteristic nonlocal length is presented. It is indicated when
the characteristic nonlocal length is less than some critical
value, such maximal transverse velocity is a constant and equal to
that of local gray soliton, otherwise such maximal velocity will
decrease with the increasing of the characteristic nonlocal
length.

\begin{acknowledgments}
We thank for the helpful discussion with professor Wei Hu.

This research was supported by the National Natural Science
Foundation of China (Grant No. 10674050), Specialized Research
Fund for the Doctoral Program of Higher Education (Grant No.
20060574006), and Program for Innovative Research Team of the
Higher Education in Guangdong (Grant No. 06CXTD005).
\end{acknowledgments}


　　　　　　　　　　　　　　　　　　　　　　　　　

\begin{thebibliography}{99}
\bibitem{1}
S. K. Turitsyn, Teor. Mat. Fiz. $\bf{64}$, 226, (1985).

\bibitem{2}
O. Bang, W. Krolikowski, J. Wyller and J. J. Rasmussen, Phys. Rev.
E $\bf{66}$, 046619 (2002).

\bibitem{3}
S. Skupin, O. Bang, D. Edmundson, and W. Krolikowski, Phys. Rev.
E, $\bf{73}$, 066603, (2006)

\bibitem{4}

Alexander I. Yakimenko, Yuri A. Zaliznyak, and Yuri Kivshar, Phys.
Rev. E, $\bf{71}$,  065603, (2005).

\bibitem{5}
C. Conti, M. Peccianti, G. Assanto, Phys. Rev. Lett. $\bf{91}$,
073901, (2003).

\bibitem{6}
C. Conti, M. Peccianti, G. Assanto, Phys. Rev. Lett. $\bf{92}$,
113902, (2004).

\bibitem{7}
W. Hu, T. Zhang, Q. Guo, L. Xuan, S. Lan, Appl. Phys. Lett.
$\bf{89}$, 071111, (2006).

\bibitem{8}
C. Rotschild, O. Cohen, O. Manela, M. Segev and T. Carmon, Phys.
Rev. Lett. $\bf{95}$, 213904 (2005).

\bibitem{9}
Q. Guo, B. Luo, F. Yi, S. Chi, Y. Xie, Phys. Rev. E, $\bf{69}$,
016602, (2004).

\bibitem{10}
S. Ouyang, Q. Guo, W. Hu, Phys. Rev. E, $\bf{74}$, 036622, (2006).

\bibitem{11}
S. Ouyang and Q. Guo, Phys. Rev. A, $\bf{76}$, 053833, (2007).

\bibitem{12}
A. W. Snyder and D. J. Mitchell, Science $\bf{276}$, 1538 (1997).

\bibitem{13}
Weiping Zhong, and Lin Yi, Phys. Rev. A, $\bf{75}$, 061801,
(2007).

\bibitem{14}
Daniel Buccoliero, Anton S. Desyatnikov, Wieslaw Krolikowski, and
Yuri S. Kivshar, Phys. Rev. Lett. $\bf{98}$, 053901, (2007).

\bibitem{15}
Dongmei Deng and Qi Guo, Opt. Lett, $\bf{32}$, 3206, (2007).

\bibitem{16}
Per Dalgaard Rasmussen, Ole Bang, Wieslaw Krolikowski, Phys. Rev.
E, $\bf{72}$, 066611, (2005).

\bibitem{17}
Shigen Ouyang, Wei Hu, and Qi Guo, Phys. Rev. A, $\bf{76}$,
053832, (2007).

\bibitem{18}
Wei Hu, Shigen Ouyang, Pingbao Yang, Qi Guo,and Sheng Lan, Phys.
Rev. A, $\bf{77}$, 033842, (2008).

\bibitem{19}
Wieslaw Krolikowski, Ole Bang, Phys. Rev. E, $\bf{63}$, 016610,
(2000).

\bibitem{20}
Alexander Dreischuh, Dragomir N. Neshev, Dan E. Petersen, Ole
Bang, and Wieslaw Krolikowski, Phys. Rev. Lett. $\bf{96}$, 043901,
(2006).

\bibitem{21}
Nikola I. Nikolov, Dragomir Neshev and Wieslaw Krolikowski, Ole
Bang, Jens Juul Rasmussen, Peter L. Christiansen, Opt. Lett,
$\bf{29}$, 286, (2004).

\bibitem{22}
Mark J. Ablowitz, Ziad H. Musslimani, Opt. Lett, $\bf{30}$, 2140,
(2005).

\bibitem{23}
G.P. Agrawal, \textsl{Nonlinear Fiber Optics}, New York:
Academeic, 1995

\bibitem{24}
Yaroslav V. Kartashov and Lluis Torner, Opt. Lett, $\bf{32}$, 946,
(2007)

\end{thebibliography}
\end{document}